\documentstyle[preprint,aps]{revtex}

\begin{document}

\draft


\title{Thermodynamics of Black Holes in Brans-Dicke Gravity}

\author{Hongsu Kim}

\address{Department of Physics\\
Sogang University, C.P.O. Box 1142, Seoul 100-611, KOREA}

\date{March, 1996}

\maketitle

\begin{abstract}
It has recently been argued that non-trivial Brans-Dicke black hole
solutions different from the usual Schwarzschild solution could exit.
We attempt here to ``censor" these non-trivial Brans-Dicke black hole
solutions by examining their thermodynamic properties.
Quantities like Hawking temperature and entropy of the black holes
are computed. Analysis of the behaviors of these thermodynamic 
quantities appears to show that even in Brans-Dicke gravity, the usual
Schwarzschild spacetime turns out to be the only physically relevant
uncharged static black hole solution.

\end{abstract}


\narrowtext


\centerline {\rm \bf 1. Introduction}

Brans-Dicke (BD) gravity [1] is perhaps the most well-known alternative 
theory of classical gravity to Einstein's general relativity. This theory
can be regarded as an economic modification of general relativity which
accomodates both Mach's principle and Dirac's large number hypothesis as
new ingredients. Ever since it first appeared, it has remained as a viable
theory of classical gravity in that it passed all the available 
observational/experimental tests provided a certain restriction on the
generic parameter, ``$\omega$" of the theory is imposed [4].
Shortly after the appearance of their first work [1], one of the authors,
C. Brans provided static, spherically-symmetric metric and scalar field
solutions to the vacuum BD field equations [2]. Since the gravitational 
collapse and the subsequent black hole formation is generally of great
interest in classical gravity, in the present work we would like to address
questions like ; under what circumstances Brans' solutions can describe
black hole spacetimes and if they actually do, they could really be
non-trivial ones different from the general relativistic black holes ?
It will then be discussed that although non-trivial BD black hole solutions
different from the usual Schwarzschild solution appears to exit as
suggested recently by Campanelli and Lousto (CL) [3], when ``censored"
by quantum aspects of black holes, namely their thermodynamics, it
can be shown that they cannot really arise in nature.
Brans [2] actually has provided exact static and isotropic solution to 
the vacuum BD field equations in four possible forms depending on the
values of the arbitrary constants appearing in the solution. In the present
work, we consider only the Brans `type I' solution since it is the only
form that is permitted for all values of the ``BD parameter", $\omega$
(the other three forms are allowed only for negative values of $\omega$,
 $\omega \le -3/2$).
In fact in this work, we will exclusively assume that the parameter
$\omega$ is {\it positive} since it has been prescribed so originally
in the BD theory itself [1] (namely according to Brans and Dicke, the
positive contribution of nearby matter to the spacetime-dependent
Newton's constant demands $\omega$ be positive) and it also has been
constrained so by experiments [4] (for $\omega \ge 500$, the theory is
in reasonable accord with all available experiments thus far.)
In fact, there is another crucial reason why $\omega$ has to be positive
from field theory's viewpoint. Namely, in order for the BD scalar field 
$\Phi$ to have ``canonical (positive-definite)" kinetic energy, $\omega$
needs to be positive.
\\
The vacuum BD gravity is described by the action (we shall work in the
Misner-Thorne-Wheeler sign convention)
\begin{eqnarray}
S = {1\over 16\pi}\int d^4x \sqrt{g} [\Phi R - \omega g^{\mu\nu}
{\nabla_{\mu}\Phi \nabla_{\nu}\Phi \over \Phi}]
\end{eqnarray}
and the field equations
\begin{eqnarray}
R_{\mu\nu} - {1\over 2}g_{\mu\nu}R &=& {\omega \over \Phi^2}
[\nabla_{\mu}\Phi \nabla_{\nu}\Phi - {1\over 2}g_{\mu\nu}
\nabla_{\alpha}\Phi \nabla^{\alpha}\Phi] \nonumber \\
&+& {1\over \Phi}[\nabla_{\mu}\nabla_{\nu}\Phi - g_{\mu\nu}
\nabla_{\alpha}\nabla^{\alpha}\Phi], \\
\nabla_{\alpha}\nabla^{\alpha}\Phi = 0. \nonumber
\end{eqnarray}
The Brans type I solution obtained in isotropic coordinates is given 
by [2]
\begin{eqnarray}
ds^2 = - \left({\tilde{r} - r_{0} \over \tilde{r} + r_{0}}\right)^{2(Q-\chi)}
dt^2 &+& \left(1 + {r_{0}\over \tilde{r}}\right)^4
\left({\tilde{r} - r_{0} \over \tilde{r} + r_{0}}\right)^{2(1-Q)}
[d\tilde{r}^2 + \tilde{r}^2 d\Omega^{2}_{2}], \nonumber \\
\Phi (\tilde{r}) &=& \left({\tilde{r} - r_{0} \over \tilde{r} + 
r_{0}}\right)^{\chi}.
\end{eqnarray}
\begin{eqnarray}
{\rm with} ~~~Q^2 + (1+{\omega \over 2})\chi^2 - Q\chi - 1 = 0.
\end{eqnarray}
Here, the arbitrary constants $Q,~\chi$ appearing in the solution are
subject to the constraint in eq.(4) and they are related to Brans'
original notation [2] by 
$Q = (1+c)/\lambda , ~~\chi = c/\lambda$ and to the notation of CL by
$Q = (1-n)$, $\chi = - (m+n)$.
Next, $\Phi(\tilde{r})$ denotes the ``Brans-Dicke scalar field" and the
quantity $r_{0}$, which is related to Brans' notation [2] by $r_{0}=B$
(or by $r_{0} = \tilde{r}_{0}$ to the notation of CL),
is a ``mass parameter" related to ``scalar" and ``tensor" mass by
$M_{s} = -\chi r_{0}$, $M_{t} = (2Q - \chi)r_{0}$ respectively.
The quantities $M_{s}$ and $M_{t}$ are constants and related to the
``Keplerian" (active gravitational) mass $M$ measured by a test particle 
by $M = M_{t} + M_{s}$. Like the ADM mass in general relativity, the 
tensor mass $M_{t}$ is positive-definite and decreases monotonically by
emitting gravitational radiation. The scalar mass $M_{s}$ and the Kepler
mass $M$, however, share none of these properties [5].
Also note that for parameter values $Q=1, ~~\chi=0$ (or $m = n = 0$ in the
notation of CL), 
the above exact 
solution in eq.(3) reduces to the usual Schwarzschild soluton in Einstein
gravity implying that it indeed is a particular solution of BD field
equations.
\\
Now, consider the coordinate transformation to the standard Schwarzschild
coordinates
\begin{eqnarray}
r = \tilde{r}(1 + {r_{0}\over \tilde{r}})^2  ~~~~{\rm or}
~~~~\tilde{r} = {1\over 2}[(r - 2r_{0}) + (r^2 - 4r_{0}r)^{1/2}]. \nonumber
\end{eqnarray}  
In terms of Schwarzschild coordinates, then, the Brans type I solution in 
eq.(3) now takes the form
\begin{eqnarray}
ds^2 = - (1 - {4r_{0}\over r})^{(Q-\chi)}dt^2
&+& {1\over (1 - {4r_{0}\over r})^{Q}}dr^2 
+ {r^2 \over (1 - {4r_{0}\over r})^{(Q-1)}}d\Omega^{2}_{2}, \nonumber \\
\Phi (r) &=& (1 - {4r_{0}\over r})^{\chi\over 2}.
\end{eqnarray}
\\
\centerline {\rm \bf 2. Non-trivial classical BD black hole solutions}
\\
As mentioned earlier, CL recently examined this vacuum solution and
claimed that under certain circumstances it could represent non-trivial
BD black hole solutions different from the usual Schwarzschild solution.
In what follows, we first give a brief review of their argument.
Obviously in order for this metric solution to represent a black hole
spacetime, the parameters $(Q, \chi)$ appearing in the solution
should satisfy certain conditions. One straightforward way of obtaining
such conditions is to consider under what circumstances an event horizon
forms. To do so, CL studied outgoing null geodesics and looked for a
condition under which the surface at $r = 4r_{0}$ (or in isotropic
coordinates, at $\tilde{r} = r_{0}$) could behave as an
event horizon. It turned out that it happens provided
\begin{eqnarray}
2Q - \chi > 1. \nonumber
\end{eqnarray}
A simple argument that can lead to this condition goes as follows ;
first, since the metric solution of the vacuum BD field equations above has
time translational $(t \to t + \delta t)$ 
isometry, a timelike Killing field
$\xi^{\mu} = \delta^{\mu}_{0}$
exits correspondingly. Now, as is well
known, in general if $\xi^{\mu}$ is a hypersurface
orthogonal Killing field which commutes with all others (if there are
more than one), then the
surfaces where $\xi^{\mu}\xi_{\mu} = 0$ are ``Killing horizons".
Thus for the case at hand, in order to see if a Killing horizon such
as an event horizon develops, we need to find out under what condition
$\xi^{\mu}\xi_{\mu}$ can have zeros. It then is
straightforward to see that $\xi^{\mu}\xi_{\mu} = g_{00} =
(1 - 4r_{0}/r)^{(Q-\chi)} = 0$ can have a zero at $r = 4r_{0}$ provided
$(Q - \chi) > 0$. In addition, certainly we do not want to have a curvature
singularity at the surface $r = 4r_{0}$ which is a candidate for
an event horizon. Thus from 
$g_{\phi \phi} = r^2(1-4r_{0}/r)^{(1 - Q)}\sin^2 \theta$,
we demand $(Q - 1) > 1$.
These two conditions, when properly put
together, then yields the above condition $2Q - \chi > 1$ which may
be thought of as the condition for this metric solution in eq.(5) to
represent possibly a black hole spacetime.
\\
Next, in order for this metric solution to represent truely a
black hole spacetime, it should have a ``regular" event horizon.
As mentioned briefly above, then, we should further require that the 
curvature, for
instance, have a non-singular behavior at the null surface $r = 4r_{0}$.
Thus in this time we consider the Kretschmann curvature invariant ;
\begin{eqnarray}
I &=& R_{\alpha \beta \gamma \delta}R^{\alpha \beta \gamma \delta} \\
&=& {(4r_{0})^2\over r^6} (1 - {4r_{0}\over r})^{(2Q-4)}\{ ({2r_{0}\over
r})^2 I_{1}(Q, \chi) + 4 ({2r_{0}\over r}) I_{2}(Q, \chi) + 6
I_{3}(Q, \chi) \} \nonumber
\end{eqnarray}
where
\begin{eqnarray}
I_{1}(Q, \chi) &=& 7Q^4 + 16Q^3 + 14Q^2 + 8Q + \chi^4 + 2\chi^3 + 
6\chi^2 - 16Q^3\chi \nonumber \\
&+& 15Q^2\chi^2 - 6Q\chi^3 - 51Q^2\chi + 20Q\chi^2 - 12Q\chi + 3, \nonumber\\
I_{2}(Q, \chi) &=& - 4Q^3 - 6Q^2 + 11Q + \chi^3 - 3\chi^2 + 7Q^2\chi
- 5Q\chi^2 + 6Q\chi - 13, \nonumber \\
I_{3}(Q, \chi) &=& 2Q^2 + \chi^2 - 2Q\chi. \nonumber   
\end{eqnarray}
Note first that this curvature invariant goes to zero as $r \to \infty$
as fast as $I \to O(r^{-6})$ and for the special case of interest,
$Q = 1$, $\chi = 0$, i.e., for the Schwarzschild solution, it reduces to
$I = {48(2r_{0})^2 \over r^6} = {48M^2 \over r^6}$ (where $M = 2r_{0}$ 	
is the ADM mass for Schwarzschild solution) as it should.
Next, it is easy to see that the condition for non-singular behavior of
the curvature invariant at $r = 4r_{0}$ amounts to the constraint
\begin{eqnarray}
Q \geq 2. \nonumber
\end{eqnarray}
Finally, put them altogether, the condition for the metric solution in
eq.(5) to represent a black hole spacetime with regular event horizon
turns out to be
\begin{eqnarray}
2Q - \chi > 1 ~~~~{\rm and} ~~~~Q \geq 2. 
\end{eqnarray}
Therefore it appears that for these values of the parameters appearing
in the solution, the metric in eq.(5) could represent a non-trivial
black hole spacetime different from the usual Schwarzschild solution
in general relativity. At this point let us recall the well-known
Hawking's theorem on black holes in BD theory [11]. Long ago, Hawking
put forward a theorem which states that stationary black holes in BD
theory are identical to those in general relativity. To be more concrete,
Hawking extended some of his theorems for general relativistic black holes
to BD theory and showed that any object collapsing to a black hole in
BD gravity must settle into final equilibrium state which is either
Schwarzschild or Kerr spacetime. And in doing so, he assumed that the
BD scalar field $\Phi$ satisfies the weak energy condition and is
constant outside the black hole. Now one may be puzzled. It is being
claimed that non-trivial BD black hole solutions different from the
Schwarzschild black hole might exit in apparent contradiction to the
Hawking's theorem. The possible existence of non-trivial BD black hole
solutions discussed here, however, does not really contradict Hawking's
theorem since they violate the weak energy condition on the BD
scalar field $\Phi$ that we now show below.  
\\
The energy-momentum tensor for the BD scalar field is given by
\begin{eqnarray}
T_{\mu\nu}(\Phi) = {1\over 8\pi}[{\omega \over \Phi^2}(\nabla_{\mu}\Phi
\nabla_{\nu}\Phi - {1\over 2}g_{\mu\nu}\nabla_{\alpha}\Phi \nabla^{\alpha} 
\Phi ) + {1\over \Phi}\nabla_{\mu}\nabla_{\nu}\Phi ]
\end{eqnarray}
from which one can readily compute
\begin{eqnarray}
T_{\mu\nu}(\Phi)\xi^{\mu}\xi^{\nu} = T_{00} = -{1\over 8\pi}(Q^2 - 1)
({2r_{0}\over r})^2 (1 - {4r_{0}\over r})^{2(Q-1)-\chi}.
\end{eqnarray}
Thus for $Q \geq 2$, which is the condition satisfied by regular BD
black hole solutions, $T_{\mu\nu}(\Phi)\xi^{\mu}\xi^{\nu} < 0$, namely the
weak energy condition on the BD scalar field is violated. And this is
the only means for the non-trivial BD black hole solutions to evade
Hawking's theorem.
Since they violate weak energy condition on BD scalar field, one may
simply reject them as unphysical solutions. One may, however, still 
take them seriously as CL did by keeping the viewpoint that demanding
the weak energy condition on the BD scalar field is not absolutely
compelling. In the present work, we choose to take the viewpoint of
the latter and then proceed further.
In the next section, we shall attempt to ``censor" these non-trivial
BD black hole solutions by examining their thermodynamic properties.
Our philosophy here is that not all classically allowed black holes
might be truely realistic and they should be further censored by
quantum aspects of black holes, i.e., their thermodynamics.
\\
\centerline {\rm \bf 3. Thermodynamics of the non-trivial BD black holes}
\\
We now turn to the investigation of the thermodynamics of non-trivial BD
black hole solutions discussed thus far.
The long-standing gap between thermodynamics and gravity has been essentially
bridged by Hawking and later by many authors [6] who first have shown, via
the study of quantum fields propagating on black hole background spacetimes,
that black holes do evaporate as if they were black bodies.
Namely the black hole thermodynamics is essentially associated with the
quantum aspect of the black hole physics. Therefore one may expect that
by examining thermodynamic properties of a black hole, one can, to some
extent, explore its quantum aspect. Besides, since the practical study
of black hole thermodynamics begins and ends with the evaluation of 
black hole's temperature and entropy, we shall attempt to compute them.
First, in order to obtain the Hawking temperature $T_{H}$ measured by an
observer in the asymptotic region, one needs to compute the surface gravity
$\kappa$ of the black hole and relate it to the temperature by
$T_{H} = \kappa/2\pi$ [6]. Although the identification of the black hole
temperature with $T_{H} = \kappa/2\pi$ was originally derived from studies
of linear, free quantized fields propagating in given black hole geometries
[6], it holds equally well for interacting fields of arbitrary spin and 
for general black hole spacetimes [9].
Thus our task reduces to the calculation of the surface gravity. In physical
terms, the surface gravity $\kappa$ is the force that must be exerted to 
hold a unit test mass at the horizon and it is given in a simple formula
as [10]
\begin{eqnarray} 
\kappa^2 &=& - {1\over 2} (\nabla^{\mu}\chi^{\nu})(\nabla_{\mu}\chi_{\nu}) \\
&=& - {1\over 2} g^{\alpha\beta}g_{\mu\nu} \Gamma^{\mu}_{\alpha\rho} 
\chi^{\rho}\Gamma^{\nu}_{\beta\lambda}\chi^{\lambda} \nonumber 
\end{eqnarray}
where $\chi^{\mu}$ is a Killing field of the given stationary black hole 
which is normal to the horizon and here the evaluation on the horizon is
assumed. For our static BD black hole given in eq.(5), $\chi^{\mu}$ is
just the timelike Killing field, $\chi^{\mu} = 
\xi^{\mu}$ and hence the surface gravity and
the Hawking temperature are computed to be
\begin{eqnarray}
T_{H} = {\kappa \over 2\pi} = {1\over 2\pi}(Q-\chi)({2r_{0}\over r^2})
(1-{4r_{0}\over r})^{(Q-\chi/2 -1)}\vert_{r=4r_{0}}
\end{eqnarray}
Having established the expression for the Hawking temperature $T_{H}$ of
the black hole, we next go on to find the ``local temperature".
Generally, the local (or ``Tolman redshifted") temperature $T(r)$ of an 
accelerated observer can be obtained by blueshifting the Hawking temperature
$T_{H}$ of an observer in the asymptotic region from infinity to a finite
point $r$.
In other words, according to ``Tolman relation" [7], the local temperature
$T(r)$ measured by a moving, accelerated detector is related to the Hawking
temperature $T_{H}$ measured by a detector in the asymtotic region by
\begin{eqnarray}
T(r) = (-g_{00})^{-1/2} T_{H}.
\end{eqnarray}
Thus the local temperature of our BD black hole is found to be
\begin{eqnarray}
T(r) = {1\over 2\pi} (Q-\chi)({2r_{0}\over r^2})  
(1-{4r_{0}\over r})^{(Q-\chi/2 -1)}\vert_{r=4r_{0}}\times
(1-{4r_{0}\over r})^{-{1\over 2}(Q-\chi)}
\end{eqnarray}
which behaves asymptotically as $T(r\rightarrow \infty) \rightarrow T_{H}$
as it should.
Next we turn to the computation of the black hole's entropy.
Generally speaking, the black hole entropy can be evaluated in three 
ways ; firstly, following Bekenstein-Hawking proposal [8], one can argue
{\it a priori} that the entropy of a black hole must be proportional to
the surface area of its event horizon (i.e., $S = {1\over 4}A$). 
Alternatively, knowing the Hawking temperature $T_{H}$ and chemical 
potentials (i.e., coulomb potential at the event horizon $\Phi_{H}$ for
the conserved U(1) charge $Q$ and angular velocity of the horizon
$\Omega_{H}$ for the conserved angular momentum $J$ generally), one may
integrate the 1st law of black hole thermodynamics
$T_{H}dS = dM - \Phi_{H}dQ - \Omega_{H}dJ$ to obtain the entropy [9].
Thirdly, according to Gibbons and Hawking [9], thermodynamic functions
including the black hole entropy can be computed directly from the
saddle point approximation to the gravitational partition function
(namely the generating functional analytically continued to the Euclidean
spacetime).
In the present case where we consider the static, isotropic black holes
in BD gravity theory, the expression for the Hawking temperature given
in eq.(11) renders it awkward to employ the second method which uses the
1st law of black hole thermodynamics. Nor can we naively adopt the 
Bekenstein-Hawking relation $S = {1\over 4}A$ to obtain black hole's
entropy. In fact, the relation $S = {1\over 4}A$ has been established 
essentially in the context of the Einstein gravity via the 1st law of
black hole thermodynamics which has been derived from the expression for 
the Bondi mass of a stationary black hole [10]. In the context of BD gravity,
due to the addition of the gravitational scalar degree of freedom (i.e.,
the BD scalar field $\Phi$), the expression for the mass of a hole is 
subject to a modification which, in turn, would result in another 
modification in the 1st law of thermodynamics violating the exact relation,
$S = {1\over 4}A$.
Therefore, here, as a reliable way of evaluating the entropy, we should 
resort to the method suggested by Gibbons and Hawking [9] since it is based 
on the fundamental definition of entropy.
To illustrate the procedure briefly, we begin with the vacuum Euclidean
BD gravity action and field equations
\begin{eqnarray}
I[g,\Phi] = -{1\over 16\pi}\left[\int_{Y}d^4x\sqrt{g}\left(\Phi R -
\omega g^{\mu\nu}{\nabla_{\mu}\Phi \nabla_{\nu}\Phi \over \Phi}\right) +
2\int_{\partial Y}d^3x\sqrt{h} \Phi (K - K_{0})\right]. 
\end{eqnarray}
The boundary term in Euclidean BD action $I[g,\Phi]$ above has been
determined as follows ; following Gibbons and Hawking [9], we start
with the form 
$\int_{\partial Y}d^3x\sqrt{h}B$
where $B  = -{1\over 8\pi}\Phi K + C$. Now for asymptotically flat
spacetimes where the boundary $\partial Y$ of the 4-dim. manifold $Y$
can be taken to be the product of the (analytically-continued) time axis
with a 2-sphere of large radius, i.e., $\partial Y = S^1\times S^2$, it 
is natural to choose $C$ so that $B$ vanishes for the flat spacetime 
metric $\eta_{\mu\nu}$. Thus we have
$B  = -{1\over 8\pi}\Phi (K - K_{0})$ where $K$ and
$K_{0}$ are traces of the second fundamental form of $\partial Y$
in the metric $g_{\mu\nu}$ and $\eta_{\mu\nu}$ respectively (one may
choose to take $B  = -{1\over 8\pi}(\Phi K - \Phi_{0} K_{0})$
where $\Phi_{0} = {\rm constant} = 1$. The choice of the latter, however,
leads to no substantial change in the conclusion that we shall draw
from the former.). Now the gravitational partition function and the 
entropy are given in the saddle point approximation by
\begin{eqnarray}
Z &=& \int [dg_{\mu\nu}][d\Phi] e^{-I[g,\Phi]} \simeq 
e^{-I[g^c,\Phi^c]}, \nonumber \\
S &=& \ln{Z} + \beta M \simeq -I[g^c,\Phi^c] + {M\over T_{H}} 
\end{eqnarray}
where the superscript ``$c$" denotes the saddle point of the action.
Since the saddle point of the action is nothing but the solution to
the vacuum BD field equations in eq.(2), first we have
\begin{eqnarray}
\int_{Y} d^4x \sqrt{g} \left(\Phi R -   
\omega g^{\mu\nu}{\nabla_{\mu}\Phi \nabla_{\nu}\Phi \over \Phi}\right) = 0 
\nonumber \end{eqnarray}
at $(g^c,\Phi^c)$ and thus the non-vanishing contribution to
$I[g^c,\Phi^c]$ comes only from the boundary term on $\partial Y$.
Here the computation of the term
$\int_{\partial Y}d^3x \sqrt{h} \Phi K_{0}$ is straightforward
since we know that for flat spacetime, $K_{0} = 2/r$.
Next from the definition of the second fundamental form $K$ [10]
and from the spherical symmetry of the BD black hole solution given
in eq.(5), we have $\int_{\partial Y}d^3x \sqrt{h} \Phi K =
\Phi {\partial \over \partial n} \int_{\partial Y} d^3x\sqrt{h}$
where ${\partial \over \partial n} \int_{\partial Y} d^3x\sqrt{h}$
is the derivative of the area $ \int_{\partial Y} d^3x\sqrt{h}$ of
$\partial Y$ as each point of $\partial Y$ is moved an equal distance
along the outward unit normal $n$. Then the result of the actual
calculation is
\begin{eqnarray}
I[g^c,\Phi^c] = 2\pi r_{0}\kappa^{-1}(3Q + \chi -2) 
+ O(r^2_{0}r^{-1}). \nonumber
\end{eqnarray}
Therefore the black hole entropy is found to be,
\begin{eqnarray}
S &=& [2 - (Q + 3\chi)]{r_{0}\over T_{H}} \\
&=& [{2 - (Q + 3\chi) \over (Q - \chi)}]\pi r^2 (1 - {4r_{0}\over r})^
{-(Q-\chi/2-1)}\vert_{r=4r_{0}}. \nonumber
\end{eqnarray}
where we have used $M = M_{t} + M_{s} = 2r_{0}(Q - \chi)$.
Note that the entropy $S$ is inversely proportional to the temperature
$T_{H}$ and it does not satisfy the usual Bekenstein-Hawking relation
[8], $S = {1\over 4}A = \pi r^2 (1 - {4r_{0}\over r})^{-(Q-1)}
\vert_{r=4r_{0}}$ as speculated earlier.
Although we have obtained the Hawking temperature $T_{H}$, the
local temperature $T(r)$ and the entropy $S$ of our BD black hole in
eq.(5), no definite statement concerning their behaviors can be made
unless the precise estimation of the allowed values of the combination
of parameters such as $(Q - \chi/2)$ and $(Q - \chi)$ are done.
Therefore, next we carry out the estimation of allowed values of
these parameters. In determining the allowed values of them, two
bottomline conditions are ; (i) the constraint equation in (4) that 
parameters
$Q$ and $\chi$ should satisfy must always be imposed and (ii) the condition
for the possible formation of non-trivial BD black holes given in eq.(7) 
must 
hold. First, as for the quantity $(Q - \chi/2)$, the condition (ii) gives
$(Q - \chi/2)> 1/2$ and $Q \geq 2$ whereas the condition (i) leads to
$-1\leq (Q - \chi/2)\leq 1$ which, when put together, yields
\begin{eqnarray}
{1\over 2} < (Q - {\chi \over 2}) \leq 1 ~~~~{\rm and} ~~~~Q \geq 2.
\end{eqnarray}
Next, it is now the turn of the quantity $(Q - \chi)$ and the condition
(i) leads to
$-\sqrt{{2\omega +4\over 2\omega +3}} \leq (Q - \chi) \leq
\sqrt{{2\omega +4\over 2\omega +3}}$ while the condition (ii) or equivalently
the condition obtained in eq.(17) gives
in addition to $Q \geq 2$,
$\{(Q - \chi) \leq {2\omega +4\over 2\omega +3}\} \cap \{(Q - \chi) <
{1\over 2}(1 - \sqrt{{3\over 2\omega +3}}) ~~{\rm or} ~~(Q - \chi) >
{1\over 2}(1 + \sqrt{{3\over 2\omega +3}})\}$ which, when put together,
yield 
\begin{eqnarray}
-\sqrt{{2\omega +4\over 2\omega +3}} \leq (Q - \chi) <
{1\over 2}(1 - \sqrt{{3\over 2\omega +3}}) ~~~{\rm and} ~~~Q \geq 2
\end{eqnarray}
or
\begin{eqnarray}
~~{1\over 2}(1 + \sqrt{{3\over 2\omega +3}}) <
(Q - \chi) \leq \sqrt{{2\omega +4\over 2\omega +3}} ~~~{\rm and} 
~~~Q \geq 2. \nonumber
\end{eqnarray}
Further, from the relationship between $\alpha \equiv (Q - \chi/2)$ and
$\beta \equiv (Q - \chi)$, that is,
$\alpha = (2\omega +4)^{-1}[(2\omega +3)\beta \pm \sqrt{(2\omega +4) -
(2\omega +3)\beta^2}]$, it is straightforward to see that, for parameter
values obtained in eqs.(17) and (18), the overlap $1/2<\alpha \leq 1$ and
${1\over 2}(1 + \sqrt{{3\over 2\omega +3}}) <
\beta \leq \sqrt{{2\omega +4\over 2\omega +3}}$
can happen as is manifest, for instance,  for $\alpha = 1/\sqrt{2}$,
$\beta = {1\over \sqrt{2}}(1 + \sqrt{{1\over 2\omega +3}})$ whereas the
other overlap can never happen since we assume $\omega > 0$.
In addition, consider as cases of special interest,  $Q = 1$, $\chi = 0$ and
$Q = 1$, $\chi = {2\over \omega +2}$. The first case is the
Schwarzschild solution as mentioned earlier and the second case corresponds
to a BD black hole with a singular horizon but satisfying the weak
energy condition on the BD scalar field. 
Finally, consider three cases (including two
cases of special interest just discussed)
\\
case (I) $~~~1/2 < (Q - \chi/2) <1$, 
$~~~{1\over 2}(1 + \sqrt{{3\over 2\omega +3}}) <
(Q - \chi) \leq \sqrt{{2\omega +4\over 2\omega +3}}$ ~~~and $~~~Q \geq 2$
\\
case (II) $~~~Q = 1$ and $~~~\chi = {2\over \omega +2}$
\\
case (III) $~~~Q = 1$ and $~~~\chi = 0$
\\
The behaviors of thermodynamic functions $T_{H}$, $T(r)$ and $S$ in
eqs.(11), (13) and (16) respectively are given for each of above three
cases as
\\
case (I) $~~~T_{H} = \infty$, $~~~T(r) = \infty$, $~~~S = 0$
\\
case (II) $~~~T_{H} = \infty$, $~~~T(r) = \infty$, $~~~S = 0$
\\
case (III) $~~~T_{H} = {1\over 16\pi r_0}$,
$~~~T(r) = {1\over 16\pi r_0}\sqrt{{r\over {r-4r_0}}}$,
$~~~S = 16\pi r^2_0$ 
\\
This behavior of thermodynamic functions reveals that BD black hole
solutions with parameter values other than $Q = 1$ and $\chi = 0$
appears to be physically irrelevant and hence should be rejected as
unphysical solutions. 
Obviously this is based on the
observation that for cases (I) and (II), we have $S = 0$ and
$T_{H} = \infty$ at  {\it any stage} of the Hawking evaporation 
which is against our conventional wisdom based on the 
usual statistical mechanics.
One might want to propose different interpretation of these behaviors
of thermodynamic quantities. Namely, since an infinite value of
temperature might signal the breakdown of the semiclassical treatment
involved in typical black hole thermodynamics, it may well be that the
non-trivial BD black holes (?) could already be quantum entities which
do not admit semiclassical treatment. It appears that the breakdown of
the semiclassical approximations can be attributed either to the failure
of weak energy condition or to the singular behavior of the horizon.
Indeed the case (I) corresponds to regular black holes violating weak
energy condition on $\Phi$ whereas the case (II) represents a singular
black hole satisfying weak energy condition. Generally, it seems that
the non-singular behavior (which requires $Q \geq 2$) and the weak 
energy condition on $\Phi$ (which demands $-1 \leq Q \leq 1$) cannot 
be simultaneously accomodated by any non-trivial BD black hole.
However, since we can always retain non-singular behavior at the expense
of weak energy condition (simply by demanding $Q \geq 2$), the
breakdown can be attributed solely to the latter. Thus to summarize,
non-trivial BD black hole solutions violating the weak energy condition
on $\Phi$, that we allowed classically, turned out to be entities
which demand more rigorous quantum treatment. Now we seem to have two
options ; first, semiclassical analysis in the black hole thermodynamics
indeed works and hence the non-trivial BD black holes should be rejected
as they fail this quantum censorship. Second, they are really quantum
entities to which the conventional black hole thermodynamics cannot
be naively applied in the first place. In the absence of a consistent
theory of quantum gravity, the second option is well beyond our scope
and in this option we do not even know whether or not these non-trivial
BD solutions can be identified with black holes at all. Thus in the
present work, we choose to take a more conservative viewpoint, namely
the first option. 
\\
\centerline {\rm \bf 4. Discussions}
\\
To conclude, in the present work we examined the thermodynamic properties of
all classically allowed  static, spherically-symmetric black hole 
solutions in 
vacuum BD theory. 
And the lesson we learned from our analysis is as
follows ; not all classically allowed  black hole solutions may be 
physically realistic. They should be further censored by quantum 
nature of black holes, namely their Hawking evaporation mechanism.
And when applied to a particular case discussed in the present work,
it turned out that the non-trivial solutions that violate the weak
energy condition on the BD scalar field and thus have been discounted in 
Hawking's theorem fail to survive the quantum censorship and hence 
would not really arise in nature. After all, stationary black holes 
in BD theory appear to be identical to those in general relativity
once they settle down.

\vspace{2cm}

{\bf \large References}

\begin{description}

\item {[1]} C. Brans and R. H. Dicke, Phys. Rev. {\bf 124}, 925 (1961). 
\item {[2]} C. Brans, Phys. Rev. {\bf 125}, 2194 (1962).
\item {[3]} M. Campanelli and C. O. Lousto, Int. J. Mod. Phys. {\bf D2},
451 (1993).
\item {[4]} See for example, C. M. Will, {\it Theory and Experiment in
Gravitational Physics}, Revised Edition (Cambridge : Cambridge University
Press, 1993).
\item {[5]} D. L. Lee, Phys. Rev. {\bf D10}, 2374 (1974).
\item {[6]} S. W. Hawking, Commun. Math. Phys. {\bf 43}, 199 (1975) ;
J. B. Hartle and S. W. Hawking, Phys. Rev. {\bf D13}, 2188 (1976) ;
R. M. Wald, Commun. Math. Phys. {\bf 45}, 9 (1975).
\item {[7]} R. C. Tolman, {\it Relativity, Thermodynamics and 
Cosmology}, Oxford, UK (1931).
\item {[8]} J. D. Bekenstein, Phys. Rev. {\bf D7}, 2333 (1973) ;
{\it ibid}, {\bf D9}, 3292 (1974).
\item {[9]} G. W. Gibbons and S. W. Hawking, Phys. Rev. {\bf D15},
2752 (1977).
\item {[10]} R. M. Wald, {\it General Relativity} (Univ. Chicago
Press, Chicago, 1984).
\item {[11]} S. W. Hawking, Commun. Math. Phys. {\bf 25}, 167 (1972).

\end{description}

\end{document}